\documentclass[a4paper]{jpconf}
\usepackage{graphicx,keyval,trig,url} 
\usepackage[T1]{fontenc}     
\usepackage[utf8]{inputenc}	
\usepackage{graphicx}

\usepackage{amsthm}
\usepackage{amssymb}
\usepackage{amsmath}
\usepackage{mathtools}
\usepackage[mathscr]{eucal}
\usepackage{lipsum}
\usepackage{cuted} 
\usepackage{blindtext}
\usepackage{color}
\usepackage{dsfont}

\usepackage{epstopdf}
\usepackage{bm}   

\usepackage{color}
\usepackage{xcolor}

\usepackage{multirow}
\usepackage{longtable}
\usepackage{graphicx}
\usepackage{amsmath}  
\usepackage{booktabs}
\usepackage{float}

\begin{document}
\title{Fault Detection of the Mooring system in Floating Offshore Wind Turbines based on the Wave-excited Linear Model}

\author{Yichao Liu$^1$, Alessandro Fontanella$^2$, Ping Wu$^3$, Riccardo M.G. Ferrari$^1$, Jan-Willem van Wingerden$^1$}

\address{$^1$ Delft Center for Systems and Control, Delft University of Technology, Delft, 2628~CD, The Netherlands.\\
$^2$ Mechanical Engineering Department, Politecnico di Milano, Milano, Via~La~Masa~1, 20156, Italy.\\
$^3$ Faculty of Mechanical Engineering \& Automation, Zhejiang Sci-Tech University, Hangzhou, 310018, China}

\ead{R.Ferrari@tudelft.nl}

\begin{abstract}
Floating Offshore Wind Turbines (FOWTs) are more prone to suffer from faults and failures than bottom-fixed counterparts due to the severe wind and wave loads typical of deep water sites.
In particular, mooring line faults may lead to unacceptably high operation and maintenance costs due to the limited accessibility of FOWTs.
Detecting the mooring line faults is therefore critical, but the application of Fault Detection (FD) techniques has not been investigated yet. 
In this paper, an FD scheme based on a wave-excited linear model is developed to detect in a reliable way critical mooring line faults occurring at the fairlead and anchor ends. 
To reach the goal, a linear model of the FOWT is obtained by approximating the wave radiation and incident wave forces. Based on this model, an observer is built to predict the rigid rotor and platform dynamics. The FD scheme is thus implemented by comparing the Mahalanobis Distance of the observer prediction error against a probabilistic detection threshold.
Numerical simulations in some selected fault scenarios show that the wave-excited linear model can predict the FOWT dynamics with good accuracy. Based on this, the FD scheme capabilities are demonstrated, showing that it is able to effectively detect two critical mooring line faults.
\end{abstract}

\section{Introduction}\label{sec:1}
Among the available renewable energy sources, wind has been deemed as one of the most promising forms when it comes to replace fossil fuels \cite{Decastro-2019}. 
In the campaign of wind exploitation that took part in the last years, offshore wind energy has received significant attention, thanks to the lower visual and acoustic impact and the abundance of space in coastal areas \cite{GWEC-2019}. 
As the exploitation of offshore wind energy moves from shallow to deep waters (i.e., a depth of more than 60\,m), Floating Offshore Wind Turbines (FOWTs) become a viable alternative than bottom-fixed ones \cite{Liu-2016}.
Nevertheless, the system complexity increases when deploying a wind turbine on a floating foundation. Moreover, FOWTs are subjected to increased loads due to the combined action of wind and waves, making them more prone to faults and failures \cite{Liu-2019} than the bottom-fixed counterparts.
FOWTs are generally operated at a considerable distance from the shore, which limits the accessibility and maintainability \cite{Levitt-2011}.

\begin{figure}
\centering \includegraphics[width=1\columnwidth]{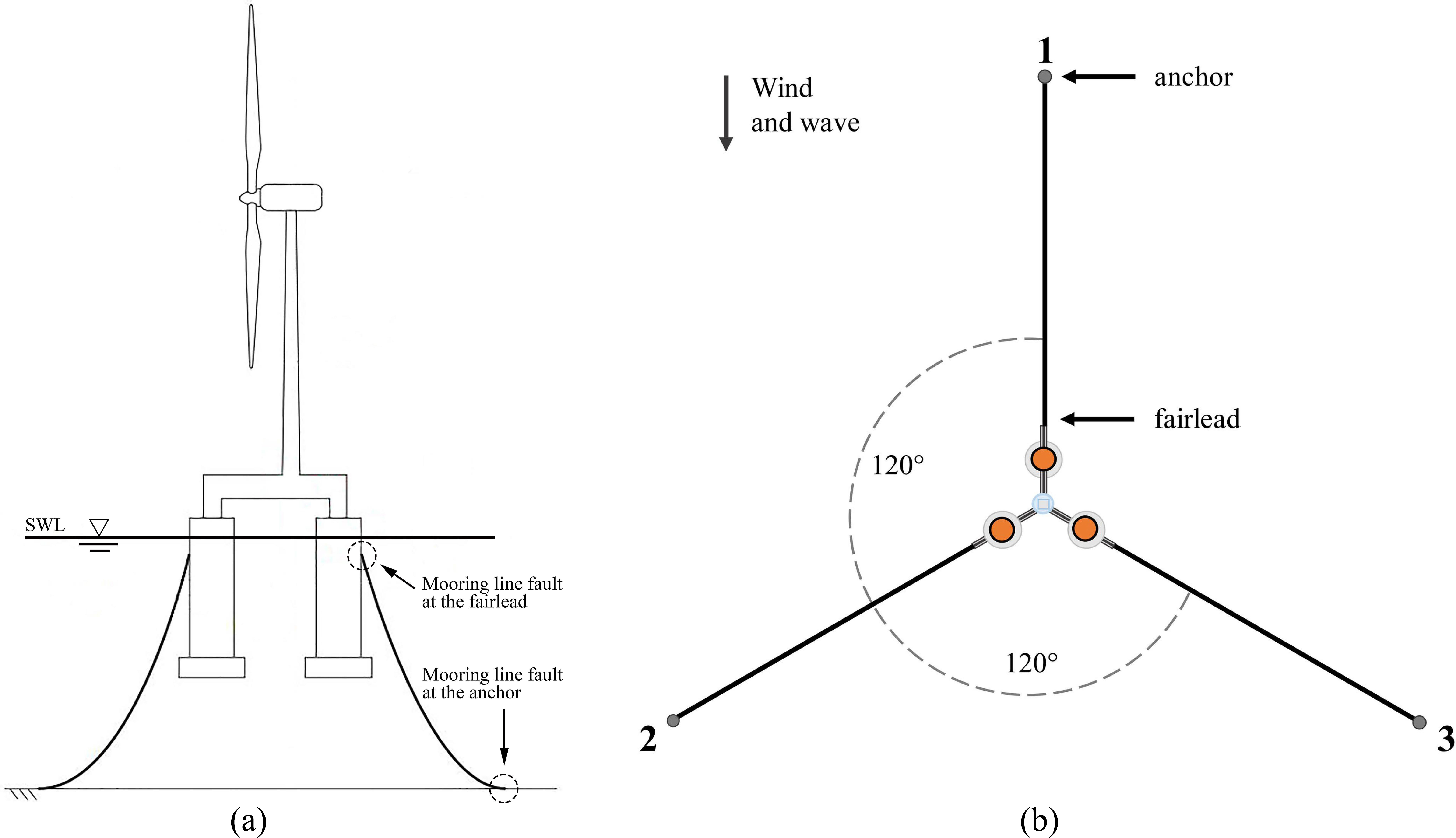}
\caption{Depicition of FOWT model. (a) DTU 10MW reference wind turbine, SWE TripleSpar floating platform and the indication of mooring line faults location. (b) Mooring line configuration.}
\label{Pic_drawing} %
\end{figure}

The mooring system of the FOWT may experience two critical kind of faults \cite{Thomas-2018} during operation: fairlead fault and anchor fault (see Fig. \ref{Pic_drawing}).
The fairlead fault consists in the failure of the top segment of the mooring line. When it occurs, the mooring line falls away and the tension to suddenly become zero. As a consequence, the FOWT drifts away from the designed location until it reaches a new equilibrium position. 
In case of anchor fault, the forces at the mooring line anchor are no more balanced by the static friction of the seabed. As a result, the anchor moves into a new equilibrium position where the unstretched length of the mooring line lying on the seabed is larger \cite{Thomas-2018}.

An effective Fault Detection (FD) scheme is of crucial importance to effectively detect the aforementioned critical mooring faults. 
Once the fault is detected by the FD information about the faulty FOWT are collected for possible fault tolerant control and subsequent condition-based maintenance. 
This will improve the reliability of FOWTs and the sustainability of the wind farm development. 

To the best of the authors' knowledge, few researches were carried out so far to investigate FD architectures for the mooring system, even though the mooring line faults of FOWTs have received increasingly significant attention.  
Bae et al. \cite{Bae_2017} investigated the dynamic response of the OC4 semi-submersible FOWT with a broken mooring line by means of numerical simulations.
More specifically, the fault scenario where one of the mooring lines suddenly disconnects from the platform (i.e. the top segment fault at the fairlead) was considered. 
Results show that in the case of the mooring line fault the FOWT undergoes a significant drift motion, which could be a risk to neighbouring wind turbines in wind farms.
In addition, Ma \textit{et al}. \cite{Ma_2019} studied the dynamic response of the same 5MW semi-submersible FOWT with a broken mooring line highlighting that a this kind of faults results into a significant change of the FOWT dynamics, and reduces the reliability of the system.

FD schemes for the mooring system, have been widely studied by the offshore oil and gas industry. Siréta and Zhang \cite{Sireta_2018} developed a fault detection system based on an artificial neural network and the measurement of the floater motion. 
Moreover, Bruggen \cite{Thomas-2018} designed a model-based FD system for the faults in a single-point mooring. 
The dynamics of the FOWT system are totally different from those of the offshore oil and gas equipment, since there is a significant influence of the complex interactions among the floating foundation, the mooring system, and the rotor blades as well as the fluid-structure interactions \cite{Wiegard_2019}.
Therefore, the existing FD schemes for the mooring system cannot be directly applied to FOWTs. 
The lack of reliable instruments for FD of the mooring system in FOWTs, together with the increased complexity of these systems with respect to bottom-fixed wind turbines and their lower maintainability, makes it urgent to develop a new FD scheme.

The present study aims to develop an effective approach to detect two critical mooring line faults in FOWTs.
To fulfill these goals, an observer that can effectively capture the global FOWT dynamics is designed based on a wave-excited linear model that includes a parametric model of radiation and wave excitation.
Based on this, mooring line faults can be detected by computing the difference between the FOWT dynamics (i.e. rotor speed and platform surge and pitch motions) as predicted by the observer and as measured from the real system. Such difference is called residual and is compared in real-time with a suitable detection threshold.
The effectiveness of the proposed model-based FD scheme is then illustrated on a case study of a 10MW FOWT. 
More specifically, it is implemented in a custom Fatigue, Aerodynamics, Structures, and Turbulence (FAST)-Simulink model. The standard FAST v8.16 distribution was modified to be able to generate, during the simulation, the critical mooring line faults described before.
The rest of the paper is formulated as follows. 
Section \ref{sec:2} elaborates the 10MW FOWT model for the FD purpose, where the fault generation of mooring line is detailed.
In section \ref{sec:3}, the overall structure of the model-based FD architecture is introduced. 
A case study employing the proposed method to detect the faults of the 10MW FOWT model is presented in section \ref{sec:4}. 
Section \ref{sec:5} draws the conclusions.

\section{Description of the simulation model} \label{sec:2}
The 10MW FOWT model, which would be used for developing the FD scheme, is introduced in this section.
It is based on the DTU 10MW reference wind turbine and the Triple-Spar floating platform \cite{fontanella2018linear,fontanella2018control}.
The lateral and top views of the 10MW FOWT are depicted in Fig.~\ref{Pic_drawing}, and its parameters are listed in Table~\ref{table:FOWT}.
\begin{table}
\centering
\caption{Specifications of the 10MW FOWT, where SWL stands for sea water level.\label{table:FOWT}}

\vspace{3mm}

\begin{tabular}{ll}
\hline 
Parameter     & Value            \\ \hline
\textbf{Turbine system}           \\
Rating and rated tip speed          & 10MW, 90m/s          \\
Rotor orientation, configuration  & Upwind, 3 blades            \\
Pitch control         & Variable speed, collective pitch        \\
Drivetrain      & Medium speed, multiple stage gearbox          \\
Rotor, hub diameter & 178.3m, 5.6m \\
Hub height & 119m \\
Cut-in, rated, cut-out wind speed & 4m/s, 11.4m/s, 25m/s \\
Cut-in, rated rotor speed & 6rpm, 9.6rpm \\ \hline
\textbf{Floating platform}  \\
Total height and draft & 66m, 56m \\
Distance from the tower center-line & 26m \\
Single column diameter & 15m \\
Column elevation above SWL & 10m \\
Elevation of tower base above SWL & 25m \\
Water displacement & 29497.7 m$^3$ \\ \hline
\textbf{Mooring lines} \\
Number of lines & 3 \\
Line angles from upwind direction & 0$^{\circ}$, 120$^{\circ}$, 240$^{\circ}$ \\
Anchor depth and radius & 180m, 599.98m \\
Fairleads above SWL & 8.7m \\
Fairleads radius & 47.181m \\
Line diameter & 0.18m \\
Total length & 707m \\
Mass/length in air & 594kg/m \\ \hline 
\end{tabular}
\end{table}

The FOWT dynamics can be described by the following discrete-time system
\begin{align}
\begin{cases}
 x(k+1)  \! \! \! &= A^0x(k)+\rho(x(k),u(k))+\beta(k-k_0)\times \\
& \;\;\; \phi_x(y(k),u(k),\vartheta_x)+\eta_x(x(k),u(k),k)  \\
 y(k) \! \! \! &= C^0x(k)+\eta_y(x(k),u(k),k)
\end{cases} \, ,
\label{eq:FOWT_DYNAMICS}
\end{align}
where $k$ is the discrete time index while $x\in\mathbb{R}^n$, $u\in\mathbb{R}^m$, $y\in\mathbb{R}^p$ denote the state, the controlled input and the measured output vectors, respectively. The matrix $A^0\in\mathbb{R}^{n\times{n}}$ and the vector field $\rho:\mathbb{R}^n\times\mathbb{R}^m\mapsto\mathbb{R}^n$ denote the nominal linear and nonlinear parts of the FOWT healthy dynamics while $C^0\in\mathbb{R}^{p\times{n}}$ is the nominal output matrix. 
The unavoidable modelling uncertainties and output disturbances are described by the functions $\eta_x:\mathbb{R}^n\times\mathbb{R}^m\times\mathbb{R}\mapsto\mathbb{R}^n$ and $\eta_y:\mathbb{R}^n\times\mathbb{R}^m\times\mathbb{R}\mapsto\mathbb{R}^p$.
The term $\beta(k-k_0)\times \phi_x(y(k),u(k),\vartheta_x)$ describes the changes in the state equation due to the occurrence of the mooring line faults at the unknown time index $k_0$.

Utilizing the model described above, the 10MW FOWT dynamics can be simulated in FAST-Simulink \cite{Jonkman-2005}.
In particular, two critical mooring line faults, the top segment fault at the fairlead and at the anchor as described in section \ref{sec:1}, can be modelled by means of the following fault function
\begin{equation}
\phi_{x}= \Delta \rho(x,u,\vartheta_{x}) \, ,
\label{eq:fault function}
\end{equation}
where $\vartheta_{x}$ is the tension of the broken mooring line at the fairlead in the top segment fault, or the variation of the anchor position in the bottom segment fault.
From the point of view of the custom FAST-Simulink code, the top segment fault is simulated by setting the tension of the failed mooring line to zero at fault time; for the bottom segment fault, it is not feasible to change the anchor location in the middle of the simulation. Instead, the unstretched length of the mooring line lying on the seabed, is changed to generate the fault at the fault time. 
Furthermore, an interface is developed in Simulink to specify all the fault information (e.g. fault time, magnitude, location) to be fed into the FAST simulator for the purpose of fault generation. 
By including the fault function into the source code of the MoorDyn module in FAST simulator and developing the fault information interface in Simulink, mooring line faults can be easily generated by the user. 
Based on such a development, the dynamics of the 10MW FOWT system can be simulated in the FAST numerical tool and two critical mooring line faults are generated in the middle of the simulation. 
Furthermore, the control system of the wind turbine, is implemented in Simulink, as reported in \cite{fontanella2018linear,fontanella2018control}. 

\section{The model-based fault diagnosis scheme} \label{sec:3}
\begin{figure}
\centering \includegraphics[width=1\columnwidth]{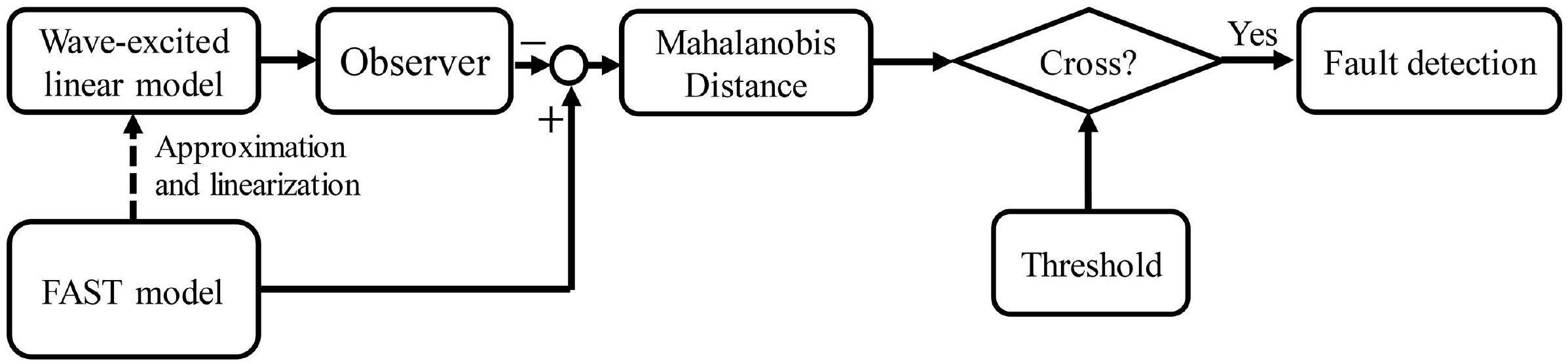}
\caption{Overall structure of the FD scheme for the mooring system in FOWTs. The wave-excited linear model is derived offline from the approximation and linearization of the FAST model while the FD scheme is implemented online.}
\label{FDscheme} %
\end{figure}

The model-based FD scheme for the FOWTs mooring system is introduced in this section. In details, it comprises an observer for the residual generation and a threshold for the fault detection.
The observer is based on a wave-excited linear model, which is an extension of the linear aero-hydro-elastic model in \cite{fontanella2018linear}. 
Based on that, the observer is implemented by means of the widely-used Kalman filter \cite{Maybeck_1990} to track the global FOWT dynamics. 
Then, the Mahalanobis Distance (MD) \cite{Maesschalck_2000} of the observer estimation error is used as a residual and compared against a probabilistic detection threshold which depends on the $\chi^2$ distribution. This allows to obtain a user defined false alarm rate, and so to guarantee robustness against stochastic uncertainties both in the model and in the measurements. 
The overall structure of the FD scheme is presented in Fig. \ref{FDscheme}.

\subsection{The Wave-excited linear model}
The wave-excited linear model consists of elastic, hydrodynamic and aerodynamic parts, for the purpose of describing the platform dynamics, the rotor dynamics around a generic steady-state configuration set by an average wind speed, and controller dynamics under a given wave condition. 
%
The model is based on the assumption that the FOWT is formed of constrained rigid bodies.
The rotor dynamics can be therefore modelled as
\begin{equation}
(J_R + \tau^2J_G)\ddot{\theta}_R = Q_{aero}-\tau Q_G \, , 
\end{equation}
where $J_R$ is the inertia of the blades and the hub, $\tau$ represents the transmission ratio between the high-speed and the low-speed shaft, $J_G$ denotes the generator inertia, $\theta$ is the rotor azimuth angle, $Q_{aero}$ and $Q_G$ denotes the aerodynamic torque and the generator torque respectively.

The floating platform motion in six Degrees of Freedom (DOFs)
$\boldsymbol{\xi}~=~(x, y, z, \rho, \beta, \sigma)^T$, namely surge, sway, heave, roll, pitch, yaw, can be calculated according to the multi-body dynamics theory \cite{Shabana_1997}
\begin{equation}
\boldsymbol{M}_{RB}{\ddot{\boldsymbol{\xi}}} + \boldsymbol{K}_{RB}{\boldsymbol{\xi}} = \boldsymbol{F} \,, 
\end{equation}
where $\boldsymbol{M}_{RB}$ and $\boldsymbol{K}_{RB}$ are the FOWT inertia and gravitational stiffness matrix. $\boldsymbol{F}$ denotes the generalized force vector for the six DOFs which contains five external forces,
\begin{equation}
\boldsymbol{F} = \boldsymbol{F}_{b} + \boldsymbol{F}_{moor} + \boldsymbol{F}_{aero} + \boldsymbol{F}_{rad} + \boldsymbol{F}_{wave} \, .
\label{eq:load}
\end{equation}
The first two terms on the right-hand side denote the restoring loads due to buoyancy and the mooring system. The rest are aerodynamic loads, hydrodynamic radiation and incident wave loads.
Actually, the first two restoring loads in equation~\eqref{eq:load} can be linearized at any static equilibrium position under the assumption that: (a) the inertia and damping of the mooring lines are neglected, (b) the relation between the platform motions and mooring forces is regarded as linear. 
The third one, responsible for the aerodynamic rotor torque and thrust, can be linearly approximated by the Taylor expansion, as reported in \cite{fontanella2018linear}.

%
%
Regarding the hydrodynamic radiation force, it is introduced in the model as
\begin{equation}
\boldsymbol{F}_{rad} \approx
-\boldsymbol{A}_{\infty}\ddot{\boldsymbol{\xi}} - \hat{\boldsymbol{\mu}}(t)
\, ,
\end{equation}
where $\boldsymbol{A}_{\infty}$ is the constant positive-definite infinite frequency added mass matrix and $\ddot{\boldsymbol{\xi}}$ the body acceleration in the six DOFs. The frequency-dependent added mass and damping associated with the fluid memory effect are approximated by $\hat{\boldsymbol{\mu}}$ that is given by a parametric linear time-invariant model in state space form as
\begin{equation}
\begin{cases} 
\dot{\boldsymbol{x}}_r = \boldsymbol{\hat{A}}_r\boldsymbol{x}_r + \boldsymbol{\hat{B}}_r\dot{\boldsymbol{\xi}} \\
\boldsymbol{\hat{\mu}} = \boldsymbol{\hat{C}}_r \boldsymbol{x}_r
\end{cases}
\, , 
\label{eq:radmodel}
\end{equation}
where the number of states in $\boldsymbol{x}_r$ corresponds to the order of the model. $\boldsymbol{\hat{A}}_r$, $\boldsymbol{\hat{B}}_r$, $\boldsymbol{\hat{C}}_r$ are the critical matrices that can be estimated by means of a frequency-domain system identification technique based on the Frequency Response Data (FRD) \cite{Ogilvie1964} as
\begin{equation}
\boldsymbol{K}(\omega) = \boldsymbol{B}(\omega)+j\omega(\boldsymbol{A}(\omega)-\boldsymbol{A}_{\infty}) \,,
\label{eq:Ogilvie}
\end{equation}
where $\boldsymbol{B}(\omega)$, $\boldsymbol{A}(\omega)$ and $\boldsymbol{A}_{\infty}$ are the frequency-dependent added damping and mass and the infinite-frequency added mass matrices.
%

In addition to the linearization of the wave radiation force, also the wave incident force $\boldsymbol{F}_{wave}$ is introduced in the model by means of a state-space linear time invariant model obtained from system identification as
\begin{equation}
\label{eq:WEM}
\begin{cases} 
\dot{\boldsymbol{x}}_w = \boldsymbol{\hat{A}}_w\boldsymbol{x}_w + \boldsymbol{\hat{B}}_w\eta \\
\boldsymbol{F}_{wave}= \boldsymbol{\hat{C}}_w \boldsymbol{x}_w 
\,, 
\end{cases}
\end{equation}
where $\eta$ represents the wave elevation at platform location,  $\boldsymbol{\hat{A}}_w$, $\boldsymbol{\hat{B}}_w$, $\boldsymbol{\hat{C}}_w$ are critical matrices and the number of states in $\boldsymbol{x}_w$ refers to the order of the parametric model.
The critical matrices are estimated by means of a time-domain system identification method to approximate the non-parametric wave force coefficients $\boldsymbol{X}(\omega)$. These are the frequency response function between wave elevation and wave forces on the platform that describe the Froude-Krylov and wave diffraction forces as:
\begin{equation}
\boldsymbol{F}_{wave} = \boldsymbol{X}(\omega)\eta(\omega)\, .
\label{eq:wavefreq}
\end{equation}

In detail, the time-domain system identification is implemented as follows:
(a) the time-domain realization  of the wave force coefficients is computed by the inverse Fourier transform of FRD $\boldsymbol{X}(\omega)$ (i.e., the impulse response);
(b) the impulse response is different from zero for $t<0$ \cite{Falnes1995379}. The impulse response is shifted forward by an arbitrary time ($t_d$) to guarantee that this is close to zero at time $t<0$. This operation ensures that the identified parametric is causal \cite{Lemmer2016290};
(c) the amplitude of the wave coefficients $\boldsymbol{X}(\omega)$ is close to zero at low frequencies. This property is obtained in the identified model by integrating the shifted impulse response in time-domain before the system identification and augmenting the resulting model with a differentiator (see \cite{Janssen2014} for an analogous discussion but for the radiation problem);
(d) the subspace identification \cite{vanderVeen_2013} is used to obtain the initial parametric linear model, which is then improved using the prediction-error minimization technique. 
\subsection{Fault detection architecture}
Based on the aforementioned linearization, all the models parts (i.e. elastic, hydrodynamic and aerodynamic), can be rewritten into a generalized state-space representation as
\begin{equation}
\begin{bmatrix} 
\ddot{\theta}_R \\
\ddot{\boldsymbol{\xi}} \\
\dot{\boldsymbol{\xi}} \\
\cmidrule(lr){1-1}
\dot{\boldsymbol{x}}_r\\
\dot{\boldsymbol{x}}_w 
\end{bmatrix} 
=
\underbrace{ \begin{bmatrix} 
\begin{array}{cc|cc}
        &         &     \boldsymbol{0} &      \boldsymbol{0}\\
      \multicolumn{2}{c|}{\smash{\raisebox{.5\normalbaselineskip}{$\boldsymbol{A}_1(U)$}}}
                   & \boldsymbol{B}_2\boldsymbol{C}_r & \boldsymbol{B}_3\boldsymbol{C}_w \\
      \hline \\[-\normalbaselineskip]
      \boldsymbol{0}&  \boldsymbol{B}_r & \boldsymbol{A}_r& \boldsymbol{0} \\
     \boldsymbol{0} &       \boldsymbol{0} & \boldsymbol{0} &  \boldsymbol{A}_w
\end{array}
\end{bmatrix} }_{\boldsymbol{A_c}}
 \begin{bmatrix} 
\dot{\theta}_R \\
\dot{\boldsymbol{\xi}} \\
\boldsymbol{\xi}\\
\cmidrule(lr){1-1}
\boldsymbol{x}_r\\
\boldsymbol{x}_w
\end{bmatrix} 
+
\underbrace{ \begin{bmatrix} 
\boldsymbol{B}_1(U) \\
\cmidrule(lr){1-1}
\boldsymbol{0} \\
\boldsymbol{0}
\end{bmatrix}  }_{\boldsymbol{B_c}}
\underbrace{ \begin{bmatrix} 
\theta \\
v \\
Q_G\\
\eta
\end{bmatrix} }_{u} \,,
\label{eq:ssmodel}
\end{equation}
%
The components of the input $u$ of the linear model are the deviations of the collective pitch angle $\theta$ and of the generator torque $Q_G$ from their steady-state values,  the horizontal turbulence $v$ and the wave elevation $\eta$. 

Then, the observer can be developed based on the linear model based on the steady-state Kalman filter \cite{Welch-1995}, which leads to
\begin{equation}
\begin{cases}
\hat{x}(k+1) \!\! &= \boldsymbol{A}\hat{x}(k)+\boldsymbol{B}u(k)+L(y(k)-\hat{y}(k)) \\
\hat{y}(k) \!\! &= \boldsymbol{C}\hat{x}(k)
\end{cases} \, ,
\label{eq:observer}
\end{equation}
where $\boldsymbol{A}$ and $\boldsymbol{B}$ are the discrete-time versions, respectively, of $\boldsymbol{A_c}$ and $\boldsymbol{B_c}$ described in equation~\eqref{eq:ssmodel}. 
$\boldsymbol{C}$ represents the output matrix with unit diagonal elements and 
$L$ is the Kalman gain matrix. 
$\hat{x}(k)$ is the state vector of the observer while $\hat{y}(k)$ represents the predicted output vector, which components include the rotor speed, the surge and pitch motions of the platform.

Based on equation~\eqref{eq:observer}, the observer prediction error $z(k)$ can be computed as
\begin{equation}
\begin{gathered}
z(k) = y(k)-\hat{y}(k)  \, ,
\end{gathered}
\label{eq:error}
\end{equation}
where $y(k)$ and $\hat{y}(k)$ denote the measurement from FAST and the observer prediction, respectively. Then the Mahalanobis Distance (MD) of the observer prediction error $z(k)$ is derived as
\begin{equation}
\begin{gathered}
d(k) = \sqrt{(z(k)-\bar{z})^{T} \sigma^{-1} (z(k)-\bar{z})}\, ,
\end{gathered}
\label{eq:r}
\end{equation}
in which $\sigma$ represents the covariance while $\bar{z}$ denotes the mean of $z$ in nominal healthy conditions.
The data set of $z$ is collected from the FAST simulations where the FOWT is operating in the nominal healthy conditions.
Consequently, $d(k)$ is regarded as the residual and compared against a probabilistic detection threshold $\bar{d}(k)$, which is designed based on the Chebychev inequality \cite{John_1984}
\begin{equation}
\begin{gathered}
 P(|d-\bar{z}|\geq \alpha S)\leq 1/\alpha^2\, ,
\end{gathered}
\label{eq:inequality}
\end{equation}
where $S$ is the standard deviation of $z$ in nominal healthy conditions. $\alpha$ is a parameter which allows to obtain a user defined expected False Alarm Rate (FAR), and so to guarantee robustness against stochastic uncertainties both in the model and in the measurements.

\section{Validation case study}
The effectiveness of the proposed model-based FD scheme is illustrated via a case study concerning the 10MW FOWT introduced in section \ref{sec:2}. The 10MW FOWT is implemented in the custom FAST-Simulink model
able to generate during the simulation the critical mooring line faults.

With regard to the model configuration, a constant wind of 16 m/s is considered in combination with an irregular wave generated according to the JONSWAP model (Significant height 2.66m, peak period 7.42s).
Each simulation lasts $1600$s at a fixed discrete time step of $0.1$s. 
The nominal healthy condition of the 10MW FOWT is simulated and the output of the FAST-Simulink model is compared to the wave-excited linear model in order to assess the prediction capability of the latter.
After this, the model-based FD scheme proposed in the present study, is implemented in Simulink. 
Four fault scenarios are considered in this paper as summarized in Table \ref{table:Fault scenarios}, all of which are generated at $1500$s for the FD purpose. The first two faults occur at the line 1 while the others appear at the line 2.
For the top segment fault, the tension of the line at the fairlead suddenly becomes zero at the fault time.
Regarding the bottom segment fault, the unstretched length of the line 1 is changed into 150m while of the line 2 is 250m at the fault time. 

\begin{table}
\centering
\caption{Fault scenarios of the mooring system in the 10MW FOWT.\label{table:Fault scenarios}}
\begin{tabular}{lll}
\hline
Load case  & Fault scenario & Parameter $\vartheta_{x}$\\ \hline
1 & Top segment fault at line 1    & 0 \\
2 & Bottom segment fault at line 1 & 150m \\
3 & Top segment fault at line 2    & 0 \\ 
4 & Bottom segment fault at line 2 & 250m \\
\hline 
\end{tabular}
\end{table}

\begin{figure}
\centering \includegraphics[width=0.8\columnwidth]{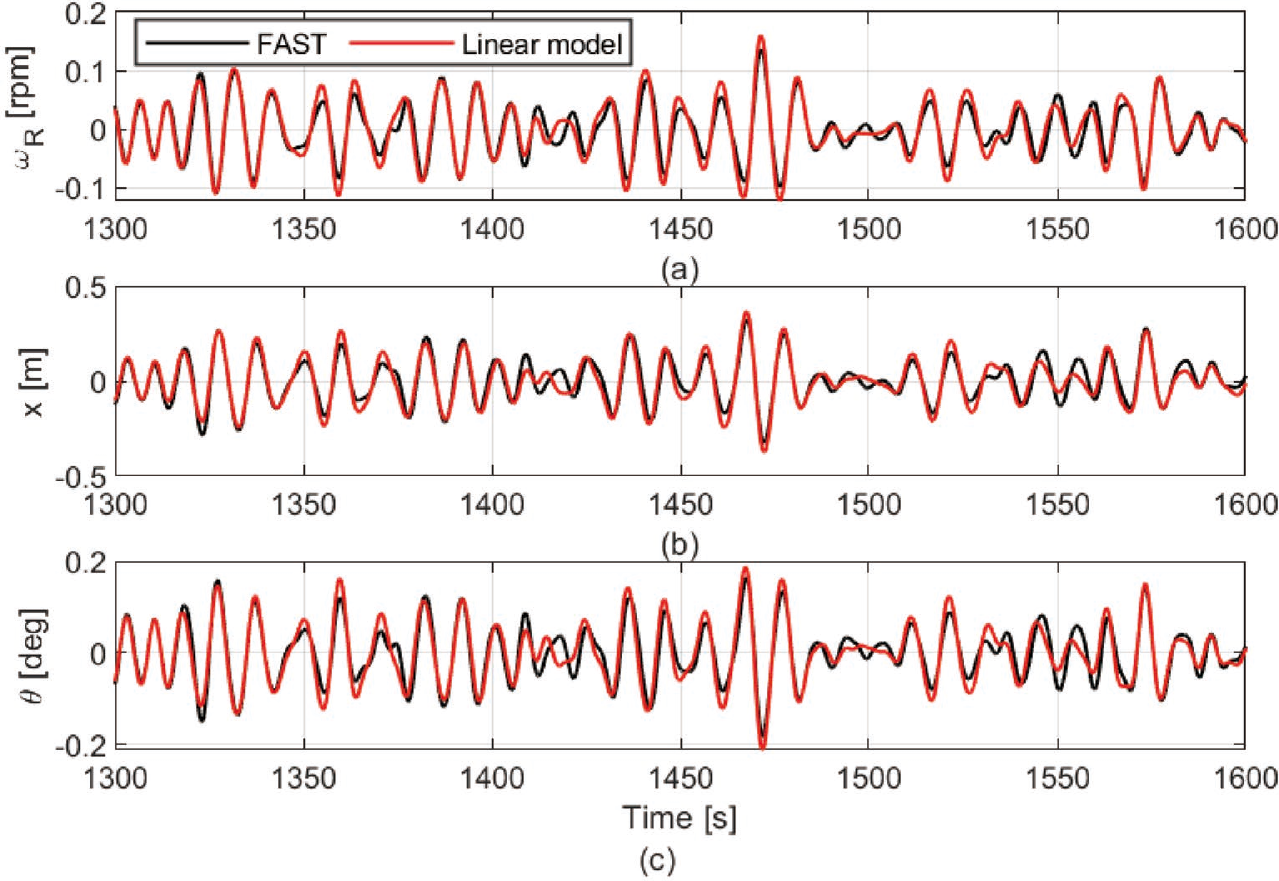}
\caption{Comparison between the wave-excited linear model and the FAST simulation results. (a) Variations of rotor speed. (b) Variations of platform surge. (c) Variations of platform pitch.}
\label{Pic_verification} %
\end{figure}

The comparison between the wave-excited linear model and the FAST simulation results is presented here for the verification purpose. 
Fig. \ref{Pic_verification} shows the variations of rotor speed, platform surge and pitch from the FAST simulation and from the wave-excited linear model. 
It can be seen that the results from the linear model, in general, match well with the output from FAST simulation.
Some deviations are observed at around 1420s and 1550s according to Fig. \ref{Pic_verification}, which are actually caused by the underlying simplifications of the aerodynamic and hydrodynamic model.
Anyway, this implies that the linear model successfully captures not only the rigid rotor dynamics but also the platform dynamics in the considered operational condition. 
Hence, the wave-excited linear model can be successfully used to develop the FD scheme for the mooring system of the 10MW FOWT. 

\begin{figure}
\centering \includegraphics[width=0.9\columnwidth]{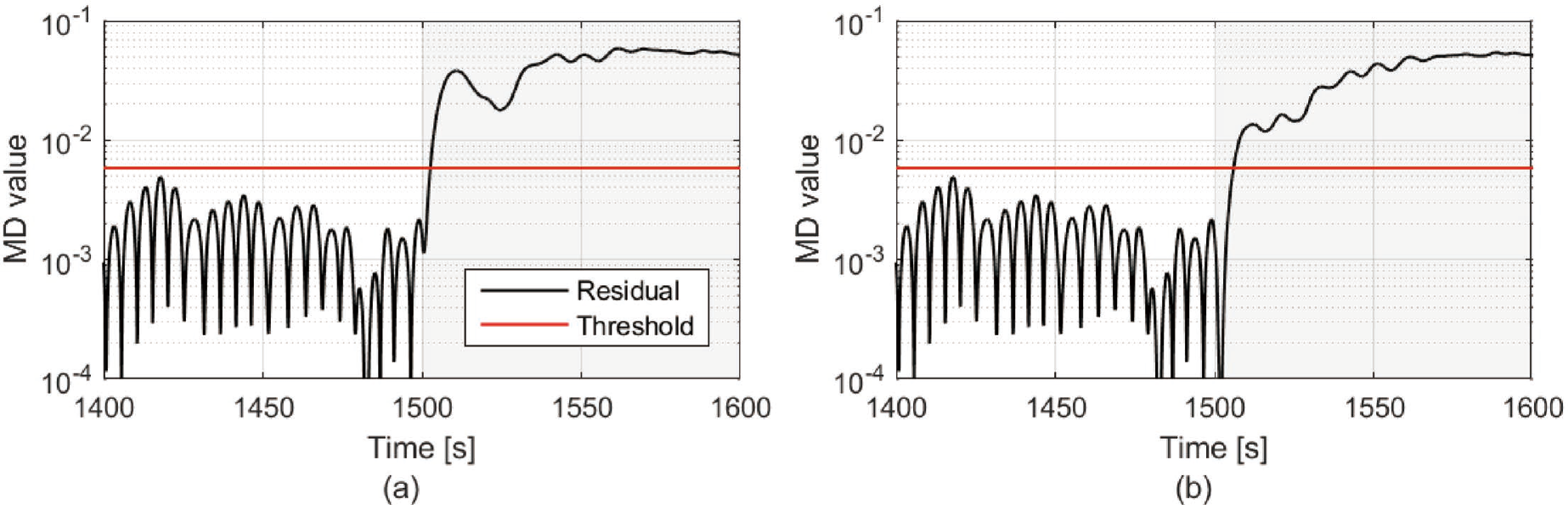}
\caption{Comparison between residuals and thresholds for fault detection of mooring line 1. (a) Top segment fault at the fairlead. (b) Bottom segment fault at the anchor. }
\label{Pic_FD} %
\end{figure}

The FD results for the mooring line 1 in first two fault scenarios are illustrated in Fig. \ref{Pic_FD}.
As visible, the proposed FD scheme is able to detect these two critical faults of the mooring line 1 effectively, since the residuals cross the corresponding thresholds immediately when faults occur. Particularly, the detection time is only 6s and 2.6s for top and bottom segment faults, which implies that the FD scheme has a quick reaction to the mooring line faults. 
Furthermore, it is noticed that the residuals in nominal healthy conditions, which may be induced by possible model and measurement uncertainties, are bounded below the MD-based threshold, which testifies the high detection capability and low false alarm of the proposed scheme.
Similar results can be found in scenarios 3-4 where the residuals cross the thresholds after 1.4s and 2.3s once the faults at the mooring line 2 appear. This implies that the faults at different mooring lines can be successfully detected as well. The plots of the results are omitted in this paper for the sake of brevity.


%

 \label{sec:4}

\section{Conclusions}
In this paper, a model-based FD scheme based on a wave-excited linear model is developed to detect two critical mooring line faults in FOWTs.
In particular, the wave-excited linear model is based on an approximated representation of the wave radiation and incident forces obtained through the system identification technique. Based on the linear model, the FD scheme is developed by comparing the MD of the observer prediction error against a probabilistic detection threshold. 

The wave-excited linear model and the proposed FD scheme are verified via a case study, concerning the 10MW FOWT model implemented in a custom Simulink-FAST model developed for the purpose.
Results have shown that: (1) the wave-excited linear model is able to capture the dynamics of the rotor and platform motions, which could be regarded as a suitable approximation of the FOWT dynamics for the FD purpose in the considered operational conditions; 
(2) the proposed FD scheme detects the top and bottom segment faults of the mooring system in short time after the fault occurs. Furthermore, this scheme is also robust against model uncertainties that do not result in false alarms, which implies that the proposed model-based FD scheme is a promising way to detect mooring line faults in FOWTs. 

Future work will include the development of fault isolation and identification algorithms. Other different conditions with wind turbulence and measurement noise will be taken into account.
 \label{sec:5}
\section{Acknowledgements}
This work has been partially funded by European Union through the Marie Sklodowska-Curie Action (Project EDOWE, grant 835901).  \label{sec:6}
\section*{References}
\bibliographystyle{unsrt} 
\bibliography{references}


\end{document}